# Predicting Gross Movie Revenue

**Sharmistha Dey**

**Date: 16th May 2016**

**Table of Content**

## 1. INTRODUCTION AND MOTIVATION

*'There is no terror in the bang, only is the anticipation of it'* – Alfred Hitchcock.

Yet there is everything in correctly anticipating the bang a movie would make in the box-office. Movies make a high profile, billion dollar industry and prediction of movie revenue can be very lucrative. Predicted revenues can be used for planning both the production and distribution stages. For example, projected gross revenue can be used to plan the remuneration of the actors and crew members as well as other parts of the budget [1].

Success or failure of a movie can depend on many factors: star-power, release date, budget, MPAA (Motion Picture Association of America) rating, plot and the highly unpredictable human reactions. The enormity of the number of exogenous variables makes manual revenue prediction process extremely difficult. However, in the era of computer and data sciences, volumes of data can be efficiently processed and modeled. Hence the tough job of predicting gross revenue of a movie can be simplified with the help of modern computing power and the historical data available as movie databases [2].

## 2. LITERATURE SURVEY

In their paper Jeffrey et al. [1] have used the publicly available data to predict the gross revenue in the USA domestic market. They attempted to predict gross revenue with the help of different sets of variables. For instance, budget, running time, star-power, MPAA ratings etc. were used to predict the movie revenue from the standpoint of production planning. Whereas the fact that approximately 25% of the gross revenue gets accumulated in the first weekend of screening; indulged them to use the first weekend collection and the number of screenings to build another more accurate prediction model. Furthermore, to find out the impact of film-critics and award nominations, they built another model using rating given by a well-known film critic and academy award nominations. There results showed that the use of opening weekend business predicts the gross revenue most accurately among all the other models.

In this paper 311 films, released in USA in the year 1998, were used to build the above mentioned linear regression models. The exhaustive list of predictor variables includes: genre, MPAA ratings, country of origin, star power, production budget, indicator variable for sequels of earlier movies, indicator variables for release during certain holiday periods of the year, number of screening in the first weekend, rating of the movie by well-known film critic and the academy award nominations.

3. DESCRIPTION OF DATA

> *'To make a great film you need three things –*
> *the script, the script and the script'* –Alfred Hitchcock

The main aim of this project is to predict the gross revenue accumulated by a movie through theater screenings. The revenue made by video or any other merchandise sale has not been considered as part of this study. The English movies released in the USA domestic market during the last six year (2010-2015) period have been chosen for this project, because that will give a large enough dataset to build and test the models.

However, only English movies released in USA domestic market with a valid MPAA rating [3] (Table 3.3) have been included in this study. The movies released briefly in 2009 December and have done most of its business in 2010 have also been considered, conversely movies briefly released in 2015 December and have done most of it business in 2016 have been excluded. Movies with any missing data (Table 3.2) have been left out as well. Furthermore, movies with less than a million dollar gross revenue have not been considered.

The online Internet Movie Database (IMDb) [2] has been used to gather the data. The raw text files were then processed with the help of a Python script developed during the data cleaning phase. It was observed that a total of 771 movies satisfied all the above constraints. Table 3.2 lists all the variables used for this project and brief descriptions of the major variables follow.

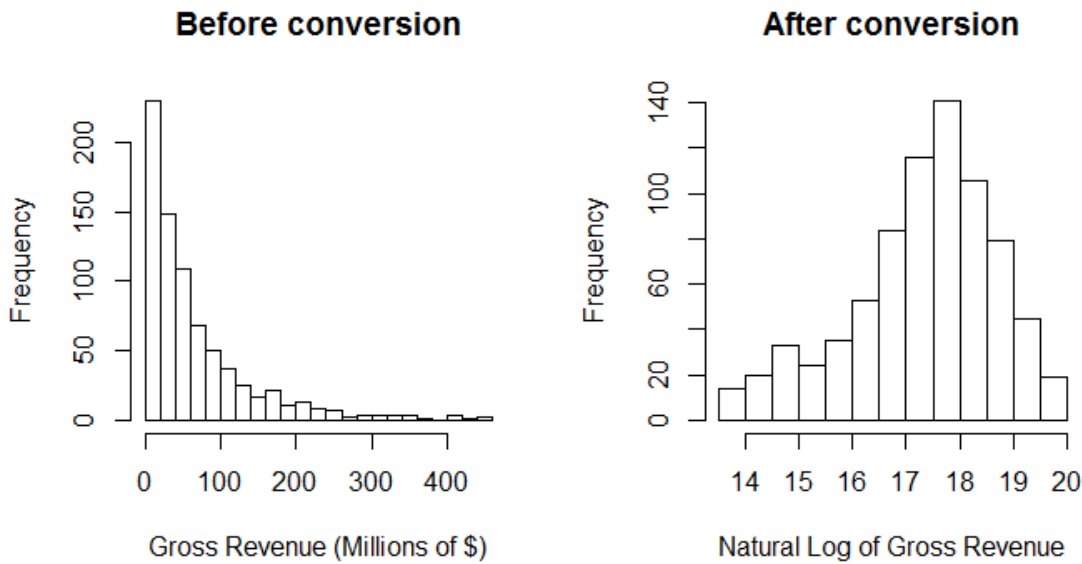

**Figure 3.1: Histogram Plot of Gross Revenue**

### 3.1. Gross-Revenue

As can be expected with any monetary data, the histogram plot of gross revenue has a long right tail (Figure 3.1). As a consequence, natural logarithm of the values has been used instead of the actual values. The other two financial information, budget and opening week revenue, have also converted into their log form before using in the models.

### 3.2. Continuous Variables

Building a linear regression model requires the explanatory variables which have a linear relation with the output variable; in other words they should have a large correlation coefficient (Table 3.1). The pair-plots in Figure 3.2 show that opening week business is highly correlated with the gross revenue (0.8) and number of screening (0.7). It is intuitive that the opening week business depends on the number of shows screened in the first week; therefore they should also have a high correlation (0.9). Abiding this logic of high correlations, the other two potential good predictors would be production-budget (0.6) and number of IMDb votes (0.5). Conversely, IMDb ratings (0.2) and running time (0.3) might not turn out to be a good choice as predictor variables.

Moreover, a visual check of Figure 3.2 confirms an asymptotic relation between number of votes and IMDb rating. This supports the intuition, since, with the increase in the number of votes, ratings start to stabilize. Besides, a large number of votes indicate higher rating, as only a good movie would gather larger audience over the time. Thus the curve slowly flattens out. However the number of votes accumulates over a period of time consequently cannot be used for prediction just after a movie releases.

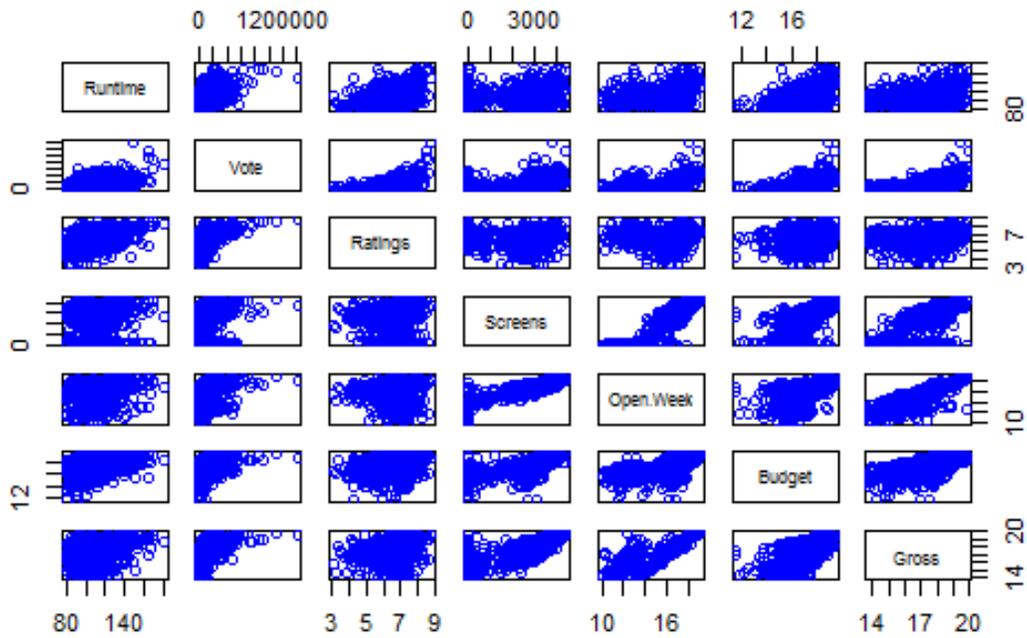

**Figure 3.2: Pair plots of the continuous variables in the movie dataset**

**Table 3.1: Correlation among Continuous qualitative Variables with Gross Revenue**

|  | **Runtime** | **Vote** | **Ratings** | **Screens** | **Open.Week** | **Budget** | **Gross** |
|---|---|---|---|---|---|---|---|
| **Runtime** | 1 | 0.479 | 0.450 | 0.085 | 0.163 | 0.370 | 0.297 |
| **Vote** |  | 1 | 0.548 | 0.283 | 0.354 | 0.436 | 0.554 |
| **Ratings** |  |  | 1 | -0.145 | -0.093 | 0.124 | 0.186 |
| **Screens** |  |  |  | 1 | 0.902 | 0.627 | 0.734 |
| **Open.Week** |  |  |  |  | 1 | 0.599 | 0.826 |
| **Budget** |  |  |  |  |  | 1 | 0.635 |
| **Gross** |  |  |  |  |  |  | 1 |

**Table 3.2: List of Variables**

| Sl No | Short Name | Long Name | Type | Value | Unit |
|---|---|---|---|---|---|
| 1 | Movie.Name | Movie Name | Character String | - | - |
| 2 | Year | Year of Release | Categorical | 2010 - 2015 | - |
| 3 | Month | Month of Release | Categorical | Jan-Dec | - |
| 4 | Sequel | If this movie is a sequel of previous movie | Binary | 0 → No<br>1 → Yes | - |
| 5 | Budget | Movie Budget | Continuous | - | USD |
| 6 | Runtime | Length of the Movie | Integer | - | Min |
| 7 | MPAA | MPAA Rating | Categorical | PG, PG-13, R | - |
| 8 | Vote | IMDB Vote | Discrete | - | Count |
| 9 | Ratings | IMDB Rating | Continuous | - | USD |
| 10 | Opening.Week | Business in the Opening Week | Continuous | - | USD |
| 11 | Screens | No of Screening in the Opening Week | Discrete | - | Count |
| 12 | Short | Genre: Short | Binary | 0 → No<br>1 → Yes | - |
| 13 | Drama | Genre: Drama | Binary | 0 → No<br>1 → Yes | - |
| 14 | Comedy | Genre: Comedy | Binary | 0 → No<br>1 → Yes | - |
| 15 | Documentary | Genre: Documentary | Binary | 0 → No<br>1 → Yes | - |
| 16 | Adult | Genre: Adult | Binary | 0 → No<br>1 → Yes | - |
| 17 | Action | Genre: Action | Binary | 0 → No<br>1 → Yes | - |
| 18 | Thriller | Genre: Thriller | Binary | 0 → No<br>1 → Yes | - |
| 19 | Romance | Genre: Romance | Binary | 0 → No<br>1 → Yes | - |
| 20 | Animation | Genre:Animation | Binary | 0 → No<br>1 → Yes | - |
| 21 | Family | Genre: Family | Binary | 0 → No<br>1 → Yes | - |
| 22 | Horror | Genre: Horror | Binary | 0 → No<br>1 → Yes | - |
| 23 | Crime | Genre: Crime | Binary | 0 → No<br>1 → Yes | - |

| Sl No | Short Name | Long Name | Type | Value | Unit |
|---|---|---|---|---|---|
| 24 | Adventure | Genre: Adventure | Binary | 0 → No<br>1 → Yes | - |
| 25 | Fantasy | Genre: Fantasy | Binary | 0 → No<br>1 → Yes | - |
| 26 | Sci.Fi | Genre: Sci-Fi | Binary | 0 → No<br>1 → Yes | - |
| 27 | Mystery | Genre: Mystery | Binary | 0 → No<br>1 → Yes | - |
| 28 | Biography | Genre: Biography | Binary | 0 → No<br>1 → Yes | - |
| 29 | History | Genre: History | Binary | 0 → No<br>1 → Yes | - |
| 30 | Sport | Genre: Sport | Binary | 0 → No<br>1 → Yes | - |
| 31 | Musical | Genre: Musical | Binary | 0 → No<br>1 → Yes | - |
| 32 | War | Genre: War | Binary | 0 → No<br>1 → Yes | - |
| 33 | Western | Genre: Western | Binary | 0 → No<br>1 → Yes | - |
| 34 | Film.Noir | Genre: Film-Noir | Binary | 0 → No<br>1 → Yes | - |
| 35 | Gross.Revenue | Gross Revenue the movie Made in USA box-office | Continuous | - | USD |

Table 3.3: MPAA Rating Explained

| Symbol | Meaning | Explanation |
|---|---|---|
| **G** | General Audience | All ages admitted. |
| **PG** | Parental guidance suggested | Some material may not be suitable for children. Parents urged to give "parental guidance". |
| **PG-13** | Parental guidance cautioned | Some material may be inappropriate for children under 13. Parents are urged to be cautious. |
| **R** | Restricted | Under 17 requires accompanying parent or adult guardian. Contains some adult material. |
| **NC-17** | Adult | No One 17 and Under Admitted. |

### 3.3. MPAA Ratings

MPAA rating or the viewing audience rating based on the movie content is assigned by 'Motion Picture Association of America' starting from the year 1990 [3]. The five ratings have been explained in the Table 3.3. The PG and PG-13 ratings have been most desirable ratings for a production house till date, since these two ratings do not restrict the viewing audience and consequently end up with a larger customer base than the restricted or adult rated movies. Figure 3.3 shows the box-plots between the MPAA ratings and gross revenue. This plot also support the preference explained above.

### 3.4. Sequel

Among the 771 movies considered for this project, 87 are sequels of previously released movies. As expected, Figure 3.4 clearly shows that sequels, on account of having successful prequels, have better market value than the non-sequel movies.

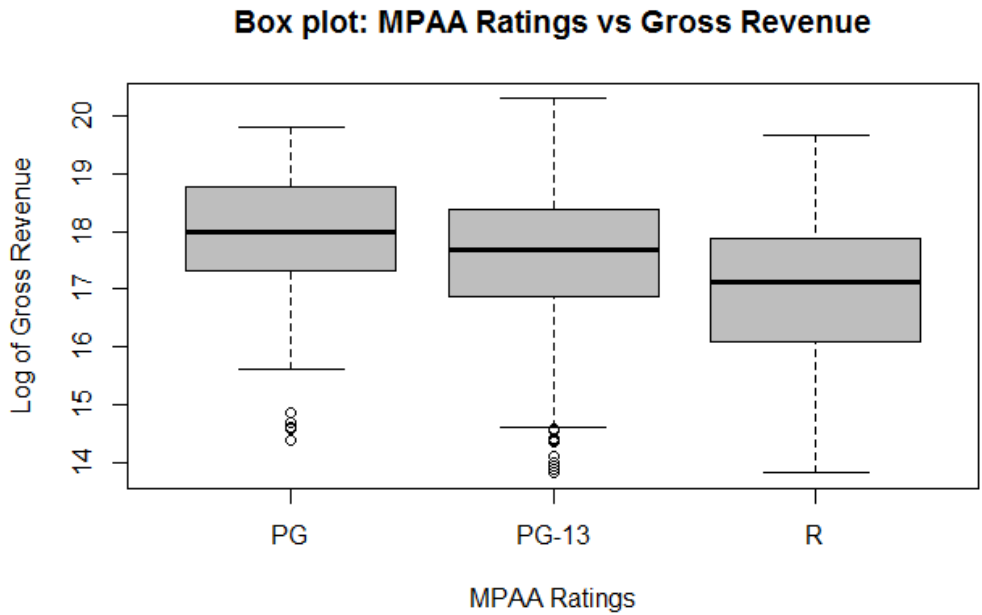

**Figure 3.3: Box-plot between the MPAA ratings and gross revenue.**

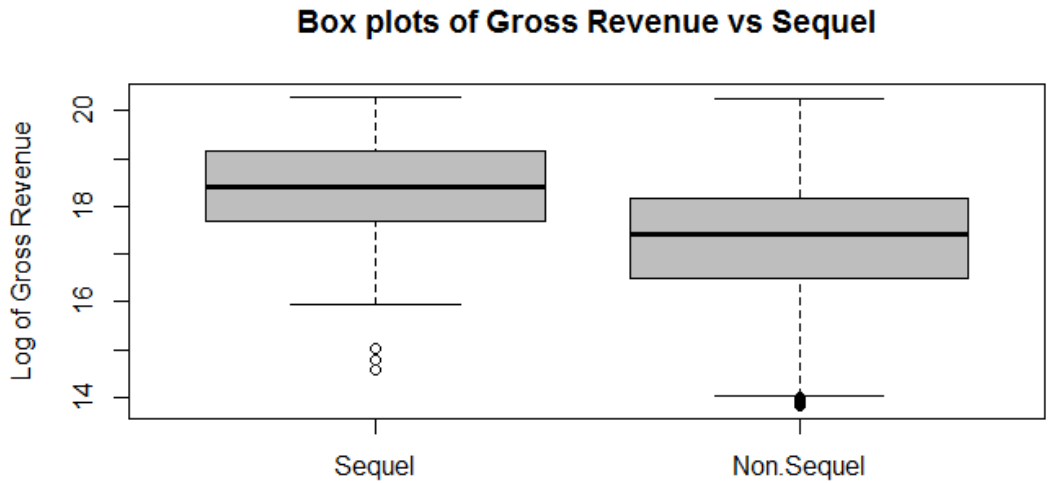

**Figure 3.4: Box-plot of Sequel vs. Gross Revenue**

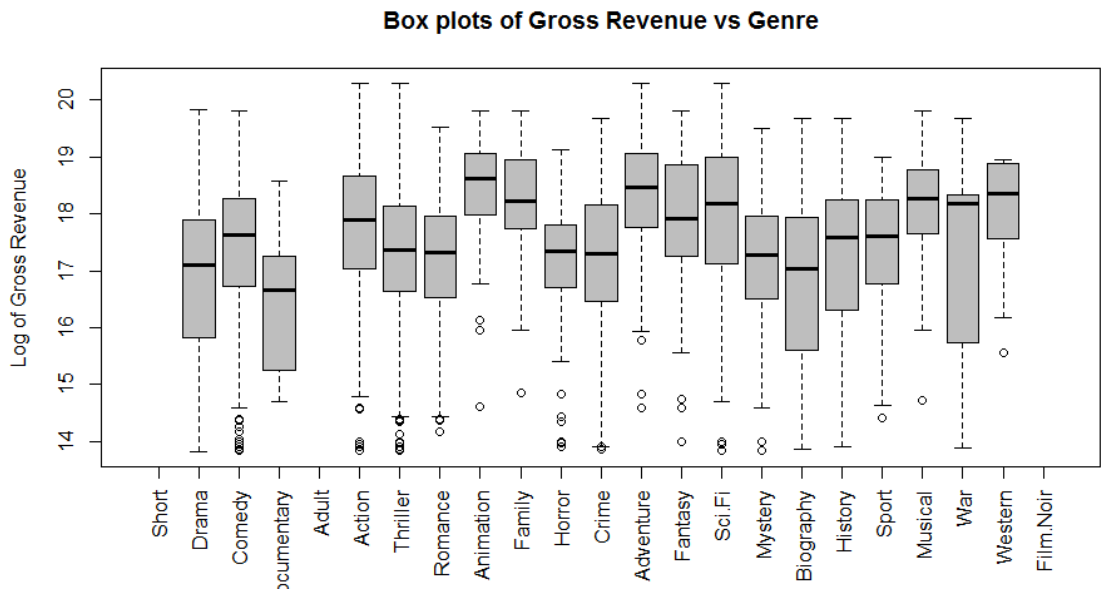

**Figure 3.5: Box plots between the different genres and the gross revenue.**

### 3.5. Genre

Genre of a movie is as important as the MPAA rating. As a case in point, majority of the time an animation movie will fall in G, PG or PG-13 category. Usually animation movies have a larger market; however these movies take longer time (sometime more than four years) to make than the other movies. Also these movies need expert technicians, who are only available with large production houses like Disney, Pixar or DreamWorks or needed to be hired on contract basis [1]. As per Figure 3.5, the other large market movie genres are family, adventure, fantasy and sci-fi. Conversely, documentary movie market is small compared to other genres.

However, many of these 23 genres are interdependent. A movie can belong to more than one genre, for example, mystery-thrillers or romantic-comedies or action-adventure etc.

To find out the latent trends among these 23 genres a factor analysis has been performed and presented later in this paper.

The 23 out of 33 predictor variables are the genre variables; however three genres Short, Adult and Film Noir were empty for the dataset under consideration. Therefore these three variables have been omitted for the rest of the paper.

## 4. SOFTWARE AND PACKAGES USED

This project was divided into three phases. The software and packages used for the respective phases have been briefly described below:

**Data Collection Phase:** Data was collected from IMDb FTP servers [2].

**Data Cleaning Phase:** In this phase a script was developed, using 'Python' scripting language, to retrieve data from the five files downloaded from IMDb [2].

**Data Visualization and Modeling Phase:** Statistical programming language 'R' was used exclusively. The packages *polycor, psych* and *caTools* in addition to the basic packages were used.

## 5. MODELING AND ANALYSIS

*"Film is one if three universal languages, the other two: mathematics and music." — Frank Capra*

The job of predicting gross revenue has been carried out for two stages of a movie production: pre-production and post-release. A short description follows -

**Pre-production:** This model forecasts the projected revenue during the production phase itself. The predictor variables used for the 'pre-production' prediction are: running time, sequel, MPAA rating, budget, month and year of release and genres.

**Post-Release:** This model projects the final revenue of a movie after its first week screening. During this phase four extra variables than Pre-Production stage are available, they are opening week revenue, number of screenings in the opening week, IMDB vote and IMDB rating.

As mentioned earlier, an exploratory factor analysis (EFA) has been performed on the genre variables before developing the linear regression models. The factors identified in the EFA stage are used in the modeling phase.

## 5.1. Factor Analysis of Binary Variables: Genre

The first step of factor analysis involves calculation of the correlation matrix of the observed variables. However, the 20 variables under consideration for EFA are not continuous variables, therefore normal way of calculating the correlation is not possible. Nevertheless, there are techniques available to do the same. Polychoric correlation co-efficient [4] is one of those techniques.

### 5.1.1. Correlation of Dichotomous (Binary) Variables: Polychoric Correlation

Ordinal variables are encountered in many fields of science. The values of an ordinal variable, or ordered categorical variables, can be arranged in a sequence, but cannot be added or multiplied. A special case of these variables are dichotomous or binary variables, which can take only two values, for example 0 and 1. The most well-known descriptive representation of a pair of ordinal variables is contingency table.

Measure of association of two ordinal variables has been an interesting study from the beginning of modern statistics. The first and probably the most popular measure is Polychoric correlation. It was first proposed by Karl Pearson in the year 1900.

The fundamental idea of Polychoric correlation measure is to assume an underlying joint normal distribution of the two ordinal variables. It is further assumed that the contingency table is the result of a double discretization of the joint distribution and the polychoric correlation coefficient is the linear correlation of the postulated joint normal distribution. Tetrachoric correlation is a special case of the polychoric correlation applicable when both observed variables are dichotomous [4].

The main disadvantage of this method is the complexity of the calculations. However, with the present improvement in computational power, the calculations are as easy as pressing a button. *'polycor'* package available in 'R' can be used to calculate the polychoric correlation among the binary genre variables. Also an in-built function *fa.poly* is available to perform factor analysis in the *'psych'* package. *fa.poly* function calculates the polychoric correlation matrix from the dichotomous data and perform the factor analysis on the correlation matrix [5].

### 5.1.2. Factor Analysis

To explore the latent trends among the 20 genres, a factor analysis has been performed. The correlations have been calculated using polychoric correlation technique and a minimal residual or ordinary least square (OLS), also known as unweighted least square method was used to find out the factor loadings. The advantages of OLS factor analysis method are ease of implementation and robustness [6]. Polychoric correlation matrices are not always positive definite, whereas OLS method does not require the positive definite criterion to be satisfied.

OLS method is a limited information method, due to the fact that polychoric correlation matrices capture only how two underlying continuous variables correlate with each other, but disregard their higher order moments like skewness and kurtosis [6].

**Table 5.1: Polychoric Correlation Co-efficient of the Binary Genre Variables**

| | Drama | Comedy | Documentary | Action | Thriller | Romance | Animation | Family | Horror | Crime | Adventure | Fantasy | Sci.Fi | Mystery | Biography | History | Sport | Musical | War | Western |
|---|---|---|---|---|---|---|---|---|---|---|---|---|---|---|---|---|---|---|---|---|
| Drama | 1.0 | -0.3 | -0.4 | -0.3 | 0.1 | 0.3 | -0.5 | -0.2 | -0.4 | 0.1 | -0.3 | -0.4 | -0.5 | 0.0 | 0.9 | 0.8 | 0.4 | 0.0 | 0.8 | -0.1 |
| Comedy | | 1.0 | 0.0 | -0.3 | -0.6 | 0.4 | 0.7 | 0.5 | -0.2 | 0.0 | 0.1 | 0.2 | -0.1 | -0.5 | -0.2 | -0.4 | -0.1 | 0.6 | -0.6 | 0.0 |
| Documentary | | | 1.0 | -0.1 | -0.3 | -0.2 | -0.1 | -0.4 | 0.0 | -0.2 | -0.5 | -0.3 | -0.3 | -0.3 | -0.2 | -0.1 | 0.0 | -0.1 | -0.1 | -0.1 |
| Action | | | | 1.0 | 0.5 | -0.3 | -0.1 | -0.2 | 0.1 | 0.2 | 0.3 | 0.0 | 0.6 | 0.2 | -0.3 | 0.0 | 0.0 | -0.6 | -0.1 | 0.3 |
| Thriller | | | | | 1.0 | -0.3 | -0.7 | -0.6 | 0.5 | 0.6 | -0.1 | -0.2 | 0.4 | 0.7 | 0.0 | 0.1 | -0.3 | -0.7 | 0.0 | 0.2 |
| Romance | | | | | | 1.0 | 0.0 | 0.1 | 0.0 | 0.0 | -0.2 | 0.0 | -0.2 | 0.0 | 0.2 | 0.0 | 0.0 | 0.3 | 0.1 | -0.5 |
| Animation | | | | | | | 1.0 | 0.9 | -0.1 | -0.5 | 0.6 | 0.6 | 0.2 | -0.4 | -0.5 | -0.5 | -0.1 | 0.6 | -0.5 | 0.2 |
| Family | | | | | | | | 1.0 | -0.2 | -0.5 | 0.7 | 0.6 | 0.1 | -0.4 | -0.2 | -0.3 | 0.1 | 0.6 | -0.2 | 0.1 |
| Horror | | | | | | | | | 1.0 | 0.0 | -0.1 | 0.4 | 0.4 | 0.7 | -0.5 | -0.5 | -0.5 | -0.3 | -0.3 | -0.2 |
| Crime | | | | | | | | | | 1.0 | -0.2 | -0.4 | 0.0 | 0.3 | 0.2 | 0.1 | -0.1 | -0.1 | -0.3 | 0.1 |
| Adventure | | | | | | | | | | | 1.0 | 0.5 | 0.4 | -0.2 | -0.3 | -0.1 | -0.1 | 0.3 | -0.1 | 0.3 |
| Fantasy | | | | | | | | | | | | 1.0 | 0.1 | 0.2 | -0.6 | -0.5 | -0.6 | 0.3 | -0.1 | 0.2 |
| Sci.Fi | | | | | | | | | | | | | 1.0 | 0.3 | -0.6 | -0.4 | 0.0 | -0.5 | -0.4 | 0.3 |
| Mystery | | | | | | | | | | | | | | 1.0 | -0.3 | -0.4 | -0.5 | -0.5 | 0.0 | 0.2 |
| Biography | | | | | | | | | | | | | | | 1.0 | 0.9 | 0.6 | 0.1 | 0.6 | -0.4 |
| History | | | | | | | | | | | | | | | | 1.0 | 0.5 | -0.1 | 0.7 | -0.3 |
| Sport | | | | | | | | | | | | | | | | | 1.0 | -0.2 | 0.4 | -0.3 |
| Musical | | | | | | | | | | | | | | | | | | 1.0 | -0.2 | -0.1 |
| War | | | | | | | | | | | | | | | | | | | 1.0 | -0.2 |
| Western | | | | | | | | | | | | | | | | | | | | 1.0 |

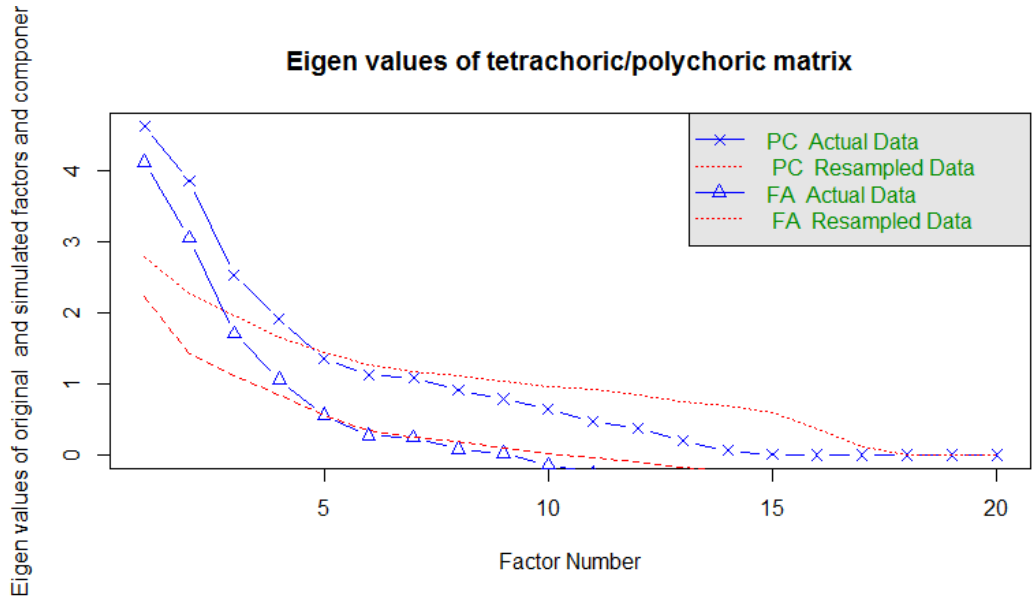

**Figure 5.1: Eigen Values of Polychoric Matrix**

The popular convention of choosing number of factors is to find the number of eigenvalues (of the correlation matrix) that are greater than one. Figure 5.1 shows that only four eigenvalues of the calculated correlation matrix are greater than one. However, a quick calculation reveals that a four-factor factor analysis (FA) explains only 64% of the total variance, while a set of eight factors explains 87.2 % of the total variance. Therefore eight factors have been considered for this project. Varimax rotation was used on the original loadings.

The loadings have been charted in Table 5.2, and a diagram showing the relations between original genre variables and the eight factors has been depicted in Figure 5.2. The solid lines in the diagram represent positive loading whereas, the dashed connections show negative ones

**Table 5.2: Factor Loadings of the 8 Factors of Genre Variables.**

| LOADINGS | MR1 | MR2 | MR3 | MR4 | MR7 | MR8 | MR6 | MR5 |
|---|---|---|---|---|---|---|---|---|
| Drama | -0.42 | 0.63 | -0.36 | -0.21 | 0.14 | 0.20 | 0.13 | -0.39 |
| Comedy | 0.34 | -0.47 | -0.34 | -0.57 | | -0.14 | -0.34 | |
| Documentary | | | | | | | 0.19 | 0.95 |
| Action | -0.12 | -0.13 | 0.82 | | -0.22 | | 0.24 | -0.14 |
| Thriller | -0.45 | | 0.36 | 0.51 | -0.52 | -0.25 | 0.10 | -0.17 |
| Romance | -0.37 | -0.32 | -0.40 | -0.46 | 0.46 | -0.30 | | -0.17 |
| Animation | 0.89 | -0.22 | | -0.24 | | | | |
| Family | 0.89 | -0.22 | | -0.18 | 0.16 | 0.14 | | |
| Horror | | -0.22 | | 0.88 | 0.16 | | -0.24 | |
| Crime | -0.40 | | | -0.11 | -0.87 | | | |
| Adventure | 0.70 | | 0.56 | -0.13 | 0.11 | -0.26 | 0.15 | -0.12 |
| Fantasy | 0.61 | -0.28 | | 0.35 | 0.30 | -0.17 | 0.16 | -0.26 |
| Sci.Fi | | -0.19 | 0.86 | | | | -0.16 | |
| Mystery | -0.34 | -0.22 | | 0.68 | -0.11 | -0.15 | 0.20 | |
| Biography | -0.18 | 0.80 | -0.31 | -0.18 | | 0.34 | | |
| History | | 0.93 | | | | | | |
| Sport | | 0.16 | | -0.12 | | 0.96 | | |
| Musical | 0.60 | | -0.44 | -0.14 | | -0.11 | 0.22 | 0.18 |
| War | -0.17 | 0.54 | | | 0.42 | | 0.44 | |
| Western | 0.12 | | | | | | 0.83 | 0.16 |
| Proportion Var | 0.19 | 0.144 | 0.129 | 0.12 | 0.084 | 0.071 | 0.068 | 0.065 |
| Cumulative Var | 0.19 | 0.334 | 0.463 | 0.583 | 0.667 | 0.738 | 0.807 | 0.872 |

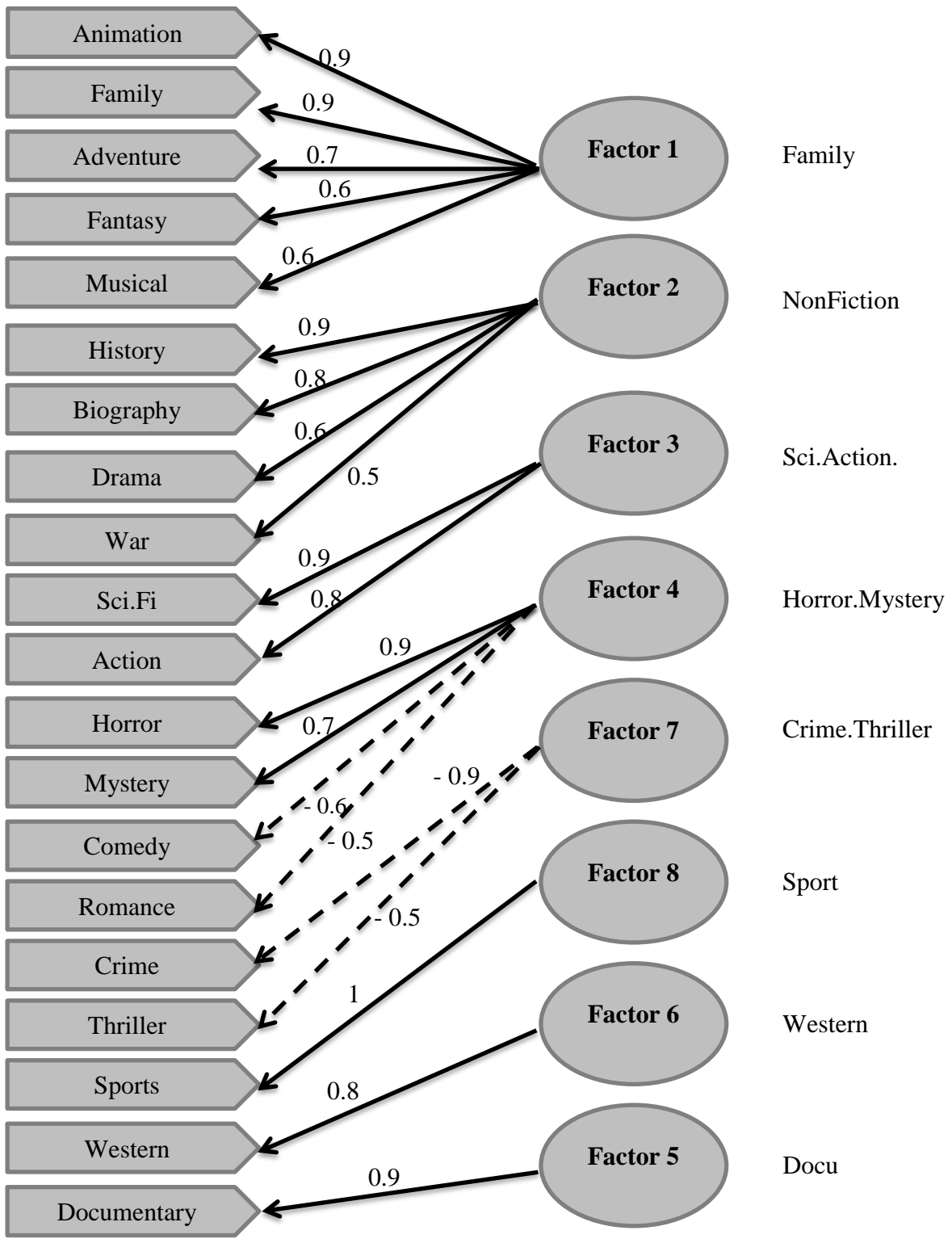

**Figure 5.2: Path Diagram to show the original Genres and Eight Factors**

## 5.2. Multiple Linear Regression Model

The main objective of this project is to predict the gross revenue of a new movie from the historical data available. The simplest and most popular parametric technique for modeling a continuous quantitative output variable with the help of multivariate inputs is *multiple linear regression* method. For this work, different combinations of all the 33 original input variables and 8 identified factors has been used to build the linear regression models. Since, prediction is the final aim, only finding the goodness of the model from the model R-square and the F-statistics is not enough. The model needed to be tested on unseen data which will represent the real world scenario [7]. To simulate that, the dataset with 769 observations has been randomly divided into three buckets: training (60%), validation (20%) and test (20%). The training set has been used to build the models, whereas the validation set data, which would not be seen by the model during training, has been used to compare the *out-of-sample* errors among the developed models. Then the best performing model is used to predict on the test dataset.

Before dividing the data into buckets, two outliers (The Avengers (2012) and Jurassic World (2015), both with more than 6 billion USD gross revenue) have been removed from the data. The final count of the observations in the dataset is 769.

### 5.2.1. Predicting From Log-Normal Distribution

Since log of the endogenous variable has been used, the predicted gross revenue in USD is required to be calculated from the output of the linear regression model. However, the fact that log-normal distributions are not symmetric (usually with a long right tail) would cause a chronic bias in the prediction if direct exponential is used.

Eq. 5.1 represents the multiple linear regression model where $b_0$ to $b_p$ are the estimated regression coefficients, $\hat{\varepsilon}_i$, the normal error with a zero mean and $\sigma^2$, the variance.

$$[\log(\hat{y}_i)]^* = b_0 + b_1 x_{1i} + b_2 x_{2i} + \cdots + b_p x_{pi} + \hat{\varepsilon}_i \qquad (Eq. 5.1)$$

The error mean and estimated variance can be represented using Eq. 5.2 and 5.3.

$$E(\hat{\varepsilon}_i) = 0 \qquad (Eq. 5.2)$$

If $s^2$ is the unbiased estimate of $\sigma^2$, then

$$s^2 = var(\hat{\varepsilon}_i) = \frac{1}{n-p} \sum_{i=1}^{n} (\log(y_i) - [\log(\hat{y}_i)]^*)^2 \qquad (Eq. 5.3)$$

In order to calculate the fitted values $\hat{y}_i$ form the log-linear model, Eq.5.4 should be used [8].

$$\hat{y}_i = \exp\left([\log(\hat{y}_i)]^* + \frac{s^2}{2}\right) \qquad (Eq. 5.4)$$

### 5.2.2. Pre-Production: Models and Prediction Accuracy

As mentioned before, prediction of revenue early in the production stage helps in planning and scheduling of a movie. In this project pre-production prediction models were built using the variables which should be known in this stage. These variables are year and month of release, MPAA rating, running time, genre, production budget and an indicator variable showing if it is a sequel. Many combinations of the input variables were evaluated, however, only the first and the best models are described here.

The first model was built with all the original inputs (Table 5.3: M1) to find out the significant predictors. Then the model was tested on the validation set (20% of the all data). However, model M1 did not perform well on the validation set and gave an *out of sample* R-Square of 0.445.

A closer look at the model M1 showed that, year and month indicator variables are not significant. The insignificance of year indicator variable can be explained using the inflation rate. The average inflation rate of USA for the last six year is 1.5 %; therefore the increment of revenue is not significantly large over the years under consideration. For the month variable, although a difference was expected for some months of the year, like the holidays in December and May, however an increase in a week might have been compensated by the business in the rest of the months.

**Table 5.3: Models and their performance on Validation and Test Data Sets**

| | Pre-Production | | | | |
|---|---|---|---|---|---|
| M1 | Year | Sequel | Budget* | 0.445 | |
| | Runtime* | MPAA* | Month | | |
| | Drama* | Comedy | Documentary | | |
| | Action | Thriller | Romance | | |
| | Animation | Family | Horror* | | |
| | Crime | Adventure | Fantasy | | |
| | Sci.Fi | Mystery | Biography* | | |
| | History | Sport | Musical | | |
| | War | Western | | | |
| M.Factor | Sequel* | Budget* | Runtime* | 0.504 | 0.468 |
| | MPAA* | Family.Factor* | SciFi.Factor | | |
| | Nonfiction.Factor* | HorrorMystery.Factor | CrimeThriller.Factor* | | |
| | Sport.Factor | Western.Factor | Docu.Factor | | |
| | Post-Release | | | | |
| Mpost.Factor | Budget* | Sequel | Family.Factor* | 0.805 | 0.771 |
| | SciFi.Factor* | Sport.Factor* | Docu.Factor* | | |
| | | Opening.Week* | Ratings* | | |

The next model (Table 5.3: M.Factor) was built using the variables: sequel, budget, runtime, MPAA rating along the eight genre factors, which were identified during factor analysis phase. The out-of-sample R-square improved to 0.504. From the Table 5.4, which charts the regression coefficients, it can be seen that some of the coefficients have negative values. This can be explained from the notion that MPAA ratings and genre-factors can have an underlying correlation. For example movies belonging to Family.Factor will also belong to PG or PG-13 categories. However, removing MPAA factor reduced the performance of the model for both the training and validation sets.

**Table 5.4: Model M.Factor developed with all the 769 Data Points**

| Item | Value | DoF |
|---|---|---|
| Residual standard error | 0.9892 | 755 |
| Multiple R-squared | 0.4729 | |
| Adjusted R-squared | 0.4638 | |
| F-statistic | 52.1 | 13 and 755 |
| p-Value | < 2.2e-16 | |

| **Coefficients:** | Estimate | Std. Error | t-Value | Pr(>|t|) |
|---|---|---|---|---|
| (Intercept) | 5.983 | 0.641 | 9.327 | <2.00E-16 |
| Sequel | 0.374 | 0.118 | 3.171 | 0.0016 |
| Budget | 0.551 | 0.040 | 13.815 | <2.00E-16 |
| Runtime | 0.016 | 0.003 | 6.067 | 2.06E-09 |
| MPAAPG-13 | 0.166 | 0.152 | 1.090 | 0.2759 |
| MPAAR | 0.006 | 0.152 | 0.037 | 0.9709 |
| Family.Factor | 0.303 | 0.068 | 4.430 | 1.08E-05 |
| Nonfiction.Factor | -0.262 | 0.050 | -5.262 | 1.86E-07 |
| SciFi.Factor | 0.007 | 0.047 | 0.152 | 0.8791 |
| HorrorMystery.Factor | 0.079 | 0.044 | 1.797 | 0.0727 |
| CrimeThriller.Factor | -0.126 | 0.043 | -2.928 | 0.0035 |
| Sport.Factor | 0.138 | 0.042 | 3.265 | 0.0012 |
| Western.Factor | -0.054 | 0.041 | -1.334 | 0.1827 |
| Docu.Factor | 0.077 | 0.040 | 1.913 | 0.0562 |

Finally the M.Fac model was marked as the best performing model for pre-production stage and then was used to predict on the test set (20 % of the all the data), which resulted an R-Square of 0.468.

**Table 5.5: Mpost.Factor, Developed with All the Data Points**

| Item | Value | DoF |
|---|---|---|
| Residual standard error | 0.6358 | 760 |
| Multiple R-squared | 0.7808 | |
| Adjusted R-squared | 0.7785 | |
| F-statistic | 338.4 | 8 and 760 |
| p-Value | < 2.2e-16 | |

| **Coefficients:** | Estimate | Std. Error | t-Value | Pr(>|t|) |
|---|---|---|---|---|
| (Intercept) | 3.60704 | 0.39556 | 9.119 | < 2.00 e-16 |
| Budget | 0.17887 | 0.02549 | 7.016 | 5.05E-12 |
| Sequel | 0.13714 | 0.07593 | 1.806 | 0.071311 |
| Family.Factor | 0.12729 | 0.02953 | 4.31 | 1.84E-05 |
| SciFi.Factor | -0.08946 | 0.02976 | -3.006 | 0.002734 |
| Sport.Factor | 0.11166 | 0.02622 | 4.259 | 2.31E-05 |
| Docu.Factor | 0.09022 | 0.02465 | 3.66 | 0.000269 |
| Opening.Week | 0.52538 | 0.01558 | 33.725 | < 2.00 e-16 |
| Ratings | 0.3588 | 0.02541 | 14.122 | < 2.00 e-16 |

**5.2.3. Post-Release: Models and Prediction Accuracy**

For the post-release prediction; budget, sequel, opening week, IMDb ratings and four out of the eight factors (Family.Factor, SciFi.Factor, Sport.Fator and Docu.Factor) have been used (Table 5.5: Mpost.Factor). This model performed far better (R-square 0.805) than the pre-production models. This was due to the fact that the opening week business and gross revenue are highly correlated. Due to the interdependency between opening week business and number of screenings; and between IMDb rating and votes, only one variable from each set have been included in the model. The IMDb rating has been chosen due to the fact the current number of votes are not representative of the immediate post release scenario. This model also performed well in the test set and resulted in an R-square of 0.771.

The two selected models for the respective stages are then built with all the 769 observations and Table 5.4 and 5.5 shows the coefficients for the same.

# 6. CONCLUSION

*"When everything gets answered, it's fake." ─ Sean Penn*

The aim of this project was to predict the gross revenue of a movie from publicly available data and by using one independent method (Factor Analysis) and one dependent method (Multiple Linear Regression). In the exploratory factor analysis (EFA) phase, eight latent factors of the twenty binary genre variables were identified to be used in the regression modeling phase.

The conclusion can be drawn from the two sets of regression models that predictions of gross movie revenue during production stage is not very accurate, however after the movie's first-week run in the theater, the projection of the final revenue becomes easier.

The models developed in the project are far from perfect. Many other variables could have been considered for the prediction process, for example, movie-plot, social media sentiment, stardom, award nominations etc. Also a more complex modeling technique like random-forest may have resulted in better predicting the gross revenue.